\documentclass[letterpaper]{article}
\usepackage{iccc}
\usepackage{longtable}
\usepackage{times}
\usepackage{helvet}
\usepackage{courier}
\usepackage{graphicx}
\usepackage{dblfloatfix}  
\usepackage{color}
\usepackage[dvipsnames]{xcolor}
\usepackage{footnote}
\usepackage{amsmath}
\usepackage{ulem}
\usepackage{url}
\usepackage{setspace}
\usepackage{makecell}
\usepackage{ascii}
\usepackage{textcomp}
\usepackage{phaistos}
\usepackage{skak,scalerel}
\definecolor{lightyellow}{RGB}{255, 255, 224}
\usepackage[skip=5pt]{caption}
\usepackage[pdftex,
            pdfauthor={Tao Long, Dorothy Zhang, Grace Li, Batool Taraif, Samia Menon, Kynnedy Simone Smith, Sitong Wang, Katy Ilonka Gero, Lydia B. Chilton},
            pdftitle={Tweetorial Hooks: Generative AI Tools to Motivate Science on Social Media},
            pdfsubject={Proceedings of the 14th Conference on Computational Creativity, ICCC '23},
            pdfkeywords={Generative AI, Science Communication, Creativity Support Tools, Social Media}]{hyperref}

\title{Tweetorial Hooks: Generative AI Tools to Motivate Science on Social Media}
\author {
    Tao Long\hspace{0.03cm}\textsuperscript{\rm \textbf{§}\scalerel*{\WhiteKingOnWhite}{Xg}}, 
    Dorothy Zhang\textsuperscript{\rm {\tiny{{\tiny{\scalerel*{\WhiteQueenOnWhite}{Xg}}}}}}, 
    Grace Li\textsuperscript{\rm {\tiny{\scalerel*{\WhiteQueenOnWhite}{Xg}}}}, 
    Batool Taraif\hspace{0.05cm}\textsuperscript{\rm {\tiny{\scalerel*{\WhiteQueenOnWhite}{Xg}}}},
    Samia Menon\textsuperscript{\rm \scalerel*{\WhiteKingOnWhite}{Xg}},\\\textbf{ {\Large 
    Kynnedy Simone Smith\textsuperscript{\rm \scalerel*{\WhiteKingOnWhite}{Xg}}, 
    Sitong Wang\textsuperscript{\rm \scalerel*{\WhiteKingOnWhite}{Xg}}, 
    Katy Ilonka Gero\textsuperscript{\rm \scalerel*{\WhiteKingOnWhite}{Xg}}, 
    Lydia B. Chilton\textsuperscript{\rm §\scalerel*{\WhiteKingOnWhite} {Xg}}}}\vspace{0.025cm}
    \\{{\Large \textsuperscript{\rm \scalerel*{\WhiteKingOnWhite}{Xg}}}\large Columbia University, New York, NY \hspace{0.6cm} 
    {\Large\textsuperscript{\rm {\tiny{\scalerel*{\WhiteQueenOnWhite}{Xg}}}}}\large Barnard College, New York, NY}\\
    {\large\textsuperscript{\textbf{§}}}{\large\fontfamily{lmtt}\selectfont{\href{mailto:long@cs.columbia.edu, chilton@cs.columbia.edu}{\{long, chilton\}@cs.columbia.edu}}}}

\begin{document} 
\maketitle
\begin{abstract}
\begin{quote}
Communicating science and technology is essential for the public to understand and engage in a rapidly changing world. Tweetorials are an emerging phenomenon where experts explain STEM topics on social media in creative and engaging ways. However, STEM experts struggle to write an engaging “hook” in the first tweet that captures the reader’s attention. We propose methods to use large language models (LLMs) to help users scaffold their process of writing a relatable hook for complex scientific topics. We demonstrate that LLMs can help writers find everyday experiences that are relatable and interesting to the public, avoid jargon, and spark curiosity. Our evaluation shows that the system reduces cognitive load and helps people write better hooks. Lastly, we discuss the importance of interactivity with LLMs to preserve the correctness, effectiveness, and authenticity of the writing.
\end{quote}
\end{abstract}

\section{Introduction}
Communicating science and technology is important for the public to understand and engage in a rapidly changing world.
Recently, a majority of the public learns about the world
not from traditional publications, but 
from social media platforms ~\cite{shearerNewsUseSocial2018}.
\textit{Tweetorials} are an emerging format for explaining complex scientific concepts on Twitter. They consist of a series of tweets that explain a technical concept in informal, narrative-driven ways~\cite{breuWhyCowCuriosity2019,breuTweetstormTweetorialsThreaded2020}. 
Whereas typical science writing is often formal, the norms of social media allow scientific conversations to take on a more personal style~\cite{bruggemann_post-normal_2020}, allowing for more creative forms of expression and engagement.

The most important part of a Tweetorial is the first tweet. This is often called a ``hook'' because it aims to hook the readers' attention and spark their curiosity so they want to read more. Although there are many ways to do this, an analysis of Tweetorial hooks~\cite{Tweetorials_tick} has shown that a common pattern is to start with a specific, relatable experience that uses no jargon. However, the challenge is to find a common experience for technical topics that a general audience of readers will find engaging.

Many STEM experts want to write creative and engaging science-related content for the public, but are not trained to do so. Their writing training is mainly for writing to peers --- other experts who are familiar with the motivation for the work, who expect expert terminologies, and who know the context surrounding the science and the formal culture of academic writing~\cite{aldousChallengesCreatingEngaging2019}. Such writing is typically (and purposefully) formulaic, and creative writing may even be discouraged in such contexts. Although there are many theories, examples, and books about public science communication, they lack mechanistic strategies proven to help people use them~\cite{howellEngagementPresentFuture2019,mcclainCriticalEvaluationScience2014,yeoPUBLICENGAGEMENTCOMMUNICATION}. Providing explicit support for informal science writing like Tweetorial hooks can better support experts in writing for the public.

We explore various ways for large language models (LLMs) to help people write engaging, creative hooks for computer science topics. 
We first explore how well LLMs can write hooks on their own by investigating three prompting strategies: instructions, instructions and examples, instructions, examples, and relatable experiences. We find that although adding examples and experiences in the prompt improves hooks, the LLMs still have much room for improvement.
Then, we design an interactive system that scaffolds the process of writing hooks but allows users to accept, reject, or improve LLM suggestions at every step. In a user study
with ten people proficient in their domain and familiar with Tweetorial hooks, we show this drastically improves their hooks and reduces cognitive load compared to writing without the system. 

\section{Related Work on LLMs and Writing Support}
Advances in LLMs have resulted in machine abilities to complete prompts with rich knowledge, commonsense reasoning, and fluent language composition~\cite{radford2019language}. Despite not being explicitly trained for specific tasks, these models possess impressive generative capabilities and can perform a diverse range of tasks. Moreover, providing just a few examples in the prompt itself can significantly enhance the quality of the model's outputs~\cite{brown_language_2020_gpt3}.

\begin{figure*}[!b]
  \centering
  \vspace{-0.3cm}
  \includegraphics[width=0.92\linewidth]{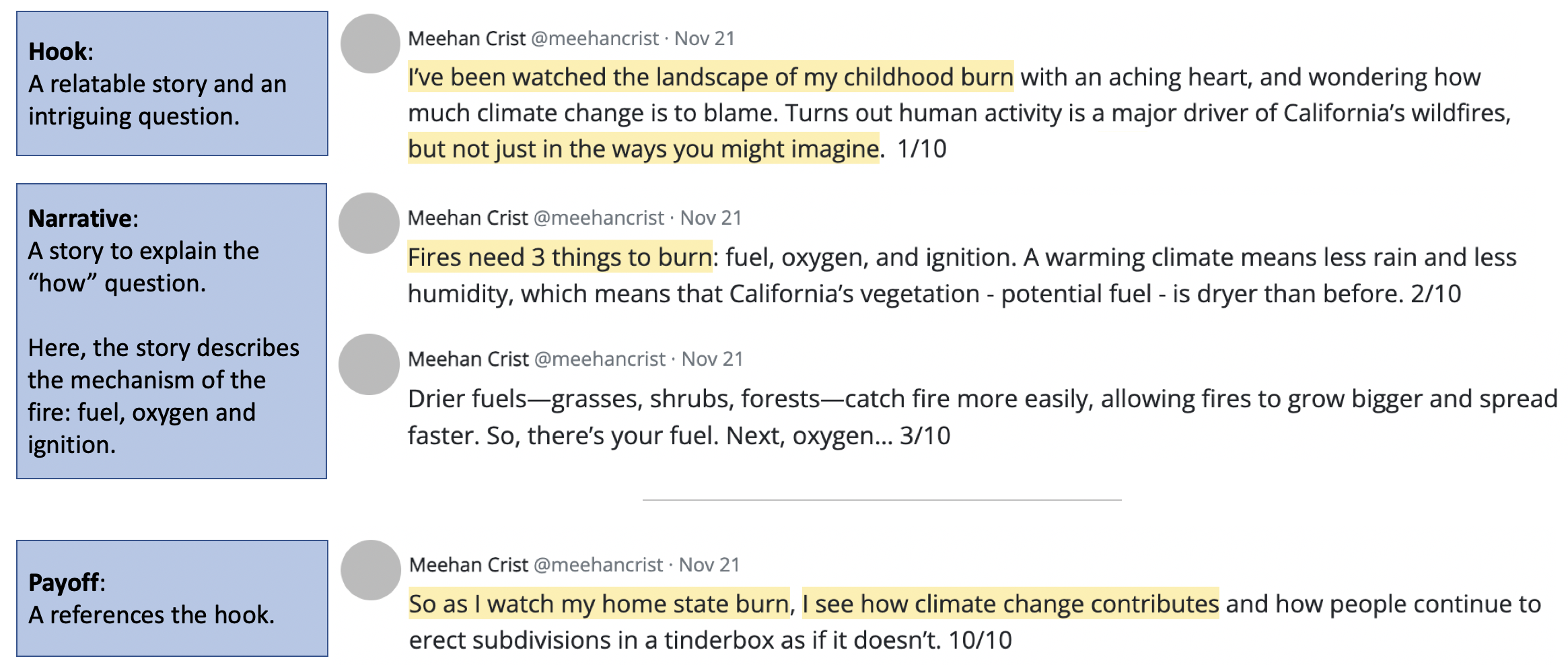}
\caption{A Tweetorial about the California wildfires \protect\cite{wildfire_tweetorial} annotated for narrative structure. Yellow highlights indicate key phrases of the hook (including the relatable detail and the intriguing question), narrative, and payoff. More annotated Tweetorial examples can be found on our website: {\small \url{http://language-play.com/tech-tweets/annotations}}}
  \vspace{-0.6cm}
\label{fig:Tweetorial1examples}
\end{figure*}

LLMs show great promise for supporting creativity and writing tasks. They can help with story writing ~\cite{calderwood_spinning,adar_TaleBrush}, brainstorming ~\cite{glassman_elephant}, and finding creative connections
~\cite{popblends} as well as story angles from press releases~\cite{Petridis2023-oh}. They have been shown to help with all three stages of the cognitive process model of writing~\cite{flower_cognitive_1981}: planning/ideation, translating/drafting, and reviewing~\cite{gero_sparks_2022}. Rather than executing these stages in a linear fashion, the writing process typically involves iterative use of these stages and requires writers to switch between their writing goals while keeping their audience in mind~\cite{emig_writing_1977}. Because of this, writing can be taxing on both the writer's short- and long-term memory, resulting in high cognitive demands~\cite{hayesnew1996}. Thus, LLMs as a writing companion and support can benefit writers in reducing mental load. 

Despite the successes of LLMs, problems remain. Language models tend to output repetitive and vague responses \cite{holtzman_curious_2020,ippolito_comparison_2019}, particularly when a prompt is underspecified or too difficult to address. One approach to address this is to chain LLM prompts together~\cite{ai_chains}: breaking down a problem into simpler and more explicit steps can make it easier for LLMs to complete. A bigger challenge is that language models have no model of truth. They learn correlations from large amounts of text, but they are not able to tell if the text they produce that includes falsehoods or offensive language~\cite{bender_dangers_2021}. Thus, LLMs may best assist writers in producing higher-quality written outputs by providing support during the writing process instead of replacing the writer and writing on their own. 

Headline writing is
an established challenge in natural language processing. Fully automated systems have some successes at generating headlines~\cite{gusev_headlines}, 
and some can even write ones in a ``clickbait'' style to hook readers \cite{jin-etal-2020-hooks,pascale_clickbait}. Although headlines do serve as hooks, traditional journalistic headlines have a different style than Tweetorial headlines. Tweetorial hooks are a little longer than headlines and use that space to start an engaging, relatable, and vivid personal anecdote. Thus the narrative, rather than the keywords, is the basis for engaging readers. This paper extends works on engaging readers with intriguing and meaningful content.

\section{Background on Tweetorial Hooks}

Tweetorials are a “collection of threaded tweets aimed at teaching users who engage with them” \cite{breuWhyCowCuriosity2019}. Across a wide range of topics from medicine, to climate science, to physics, to computer science, these tweets always introduce a technically complicated concept or answer a popular science question through  informal, narrative-driven, and creative writing ~\cite{breuTweetstormTweetorialsThreaded2020,gero_sparks_2022}. Figure \ref{fig:Tweetorial1examples} shows the first, last, and some middle tweets that form the overall narrative. Hooks are the first tweet that grab readers' attention and pull them into the narrative.

Previous work~\cite{Tweetorials_tick} has analyzed Tweetorial hooks and described 
% what they need to make them good
attributes of high-quality ones: 1) a relatable and interesting example as a lead-in and 2) an intriguing question that is driving and specific that sparks readers’ curiosity. Relatable and specific content can take many forms. 
It can relate a topic to things in the news, refute a popular misconception, or take a common daily experience and help explain it. For the language to be relatable, the hook should not include jargon. Using unfamiliar technical terminology undermines the purpose of engaging the public \cite{bullock2019jargon}. Then, an intriguing question will be directly or implicitly proposed to the readers to help spark curiosity and draw them to the following thread. The unanswered question will connect the previous relatable example to the following threads of explanations. Thus, we establish a list of requirements (R) for a relatable and engaging Tweetorial hook:

\begin{itemize}
    \item \underline{\textit{\textbf{R1} - Jargon-Free:}} Does the hook avoid jargon or unexplained terminology so the general audience can understand it easily?
    \item \underline{\textit{\textbf{R2} - Specific and Relatable Example(s):}} Does the hook include specific and relatable example(s) about the topic?
    \item \underline{\textit{\textbf{R3} - Sparks Curiosity:}} Does the hook drive readers to continue reading and satisfy their curiosity?
\end{itemize}

Here are two examples of Tweetorial hooks for computer science topics that exhibit all these properties: 
\begin{itemize}

\item Virtual Private Network (VPN): \textit{``I once torrented Last Week Tonight --- then my landlord got a complaint from Comcast! WTH? My friends never got caught. Ugh. So here are things I wished I had known about how to be sneaky on the internet:''}

\item Language Models: \textit{``My son relies on his Alexa to help with his math homework every single night. While I am concerned about his learning, I am interested in how Alexa understands what he is saying? Is it the same way that humans understand language? What is the difference? A thread on how language models helps with this:''}
  
\end{itemize}

For each hook, the topic is motivated by an everyday experience. For VPNs, the experience is torrenting. For language models, the experience is Alexa.  Each experience is told in a personal way (\textit{``then my landlord got a complaint from Comcast!''}), with informal language (``\textit{Wth? My friends never got caught. Ugh.}''). Often they have very specific details (\textit{``Last Week Tonight''}). They don't contain jargon---other than when mentioning the name of the topic towards the end of the hook. And they have a question or implied question that sparks curiosity and drives the reader to learn more (\textit{``how to be sneaky on the internet''}). This is a lot to achieve in one tweet.

Studies on Tweetorial writing have shown that writing hooks is a key challenge for STEM experts~\cite{Tweetorials_tick}. They are trained to write about their work in a formal tone for other experts, and it is difficult to go against that training. Also, they feel uncomfortable using subjective and informal language and avoid personal details, even though 80\% of the Tweetorials have them. 

In an exploratory study using LLMs to support Tweetorial writers, one of the major use cases was 
ideating concrete examples for the hook~\cite{gero_sparks_2022}. This indicates there is potential to help STEM experts write in informal styles.
We build on this potential by studying LLMs' potential to write hooks, then designing a workflow to scaffold writers' hook writing process and using LLMs to suggest options for relatable experiences that are jargon-free and can spark curiosity.

\section{Study 1: Prompt Engineering Study}
We first investigate how well an LLM can write hooks without human intervention. Then, we compare the performance of three prompting strategies and use expert annotators to evaluate the outputs. 

\subsubsection{Participants and Procedures}
We identified 30 technical computer science topics that are important for a general audience to understand. We selected them randomly from the \textit{Sideways Dictionary} \cite{sidewaysdictionary} --- a website for journalists to find accessible explanations for common technical terms. These terms included such as \textit{Database}, \textit{Browser Hijacking}, \textit{Programming Language}, \textit{Internet Service Provider}, and \textit{Autocomplete} (See \hyperref[appendix:A]{Appendix} for the complete list).

The three prompting strategies (PS) we compared are:
\begin{itemize} 
    \item \underline{\textit{\textbf{PS1} (Instructions only)}} is the most basic strategy which asks for a hook and provides simple instructions that the hook should be jargon-free, include a relatable and specific example, and spark curiosity. This is the bare minimum needed to explain to the LLM the goal of a hook.
    \item \underline{\textit{\textbf{PS2} (Examples and Instructions)}} has all the instructions from PS1, and adds five examples of good hooks we identified and collected inside the team. These hooks were taken from published Tweetorials and edited lightly for clarity. Adding examples is a known technique to help the LLM learn the ``styles'' that are difficult to describe or to phrase in specific instructions
    % . It is good for conveying 
    such as
    writing objective, writing structure, diction, and tone.
    % , etc.
    \item \underline{\textit{\textbf{PS3} (Examples, Chained User Details, and Instructions)}} is a three-stage pipeline that chains LLM prompts together \cite{ai_chains}, in addition to all the content from PS2. It first asks for the user's topic to generate everyday examples, then common experiences, then a specific personal anecdote about this experience. LLM chaining is known to work well when instructions are complicated. It breaks down the problem into simpler steps and builds up to a complex output.
\end{itemize}
% \vspace{-0.3cm}

\begin{figure}[!ht]
  \centering
  \includegraphics[width=\linewidth]{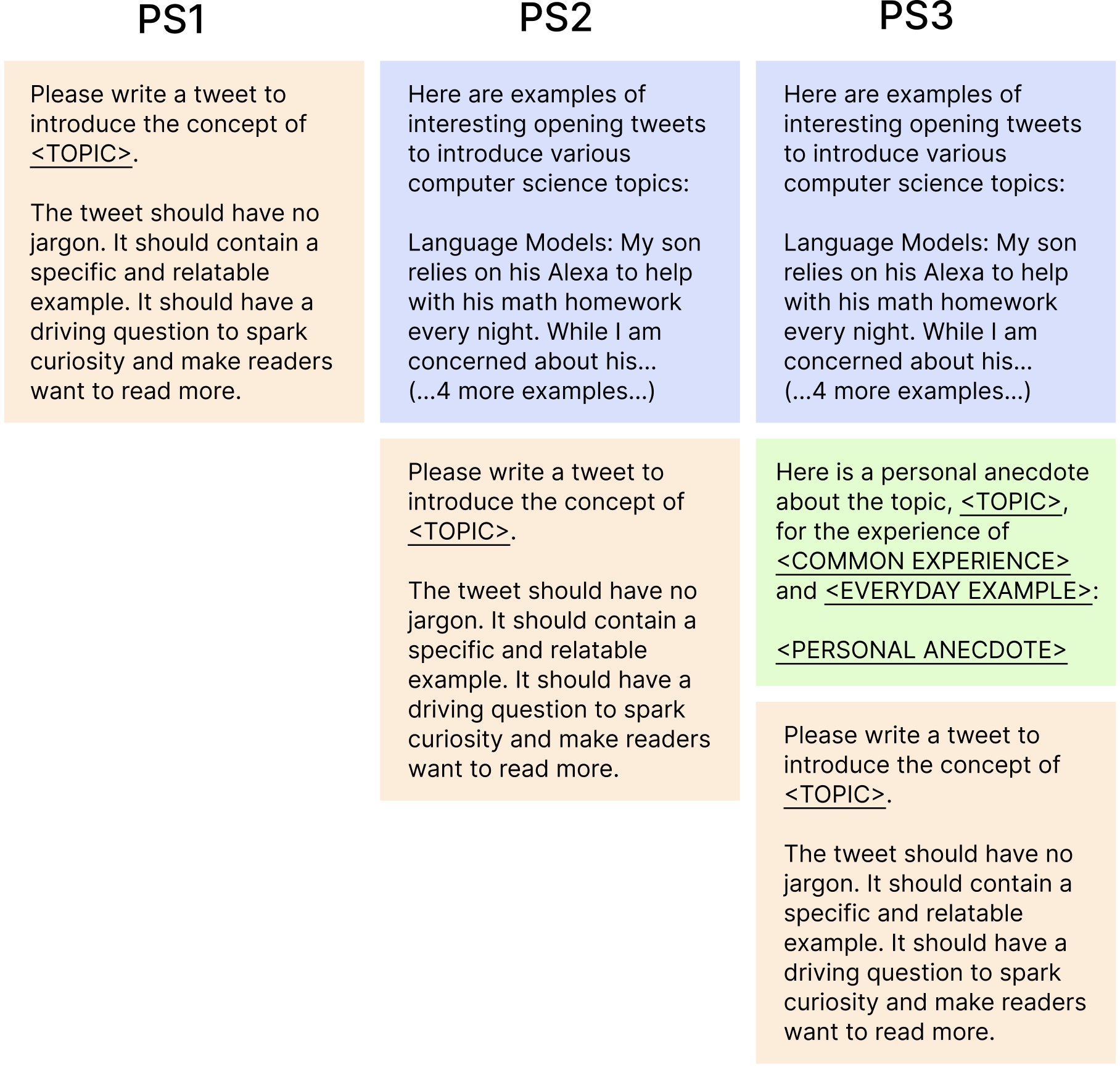}
\caption{An illustration of the three prompting strategies}
\label{fig:PromptingStrategies}
\end{figure}
% \vspace{-0.3cm}

We used OpenAI's GPT-3 API and its \textit{text-davinci-003} model with the default settings for all parameters, as it was the most capable model available at the time of our study. 

In this study, we investigate the following hypothesis:

\textbf{Hypothesis \#1: PS3 will attain the highest overall score and outperform both PS1 and PS2 across all three rubric categories.} We believe that the prompt chaining will break down the complex hook writing task into simpler steps that LLMs will be better able to solve one at a time.

To evaluate the three prompting strategies, we hired three annotators with professional training in communication and writing to judge the hooks' quality. Each annotator rated 270 hooks --- 30 topics with three prompting strategies and three generations each. The annotators were paid \$20 per hour and evaluated each hook on a 1 to 5 scale based on the criteria: whether it is jargon-free (R1), contains a relatable example (R2), and sparks curiosity (R3). They received a detailed annotation rubric with examples (See \hyperref[appendix:A]{Appendix}).

\subsection{Results}
%Report Kappa scores:
Overall, the annotators had fair agreement on their assessment, with a Fleiss’ kappa of 0.23.
%The Cohen’s kappa for inter-rater reliability was 0.24 for R1 (jargon-free), 0.27 for R2 (relatable example), and 0.18 for R3 (sparks curiosity). The Fleiss’ kappa for all three annotators is 0.23.

\begin{figure}[!ht]
  \centering
  \includegraphics[width=\linewidth]{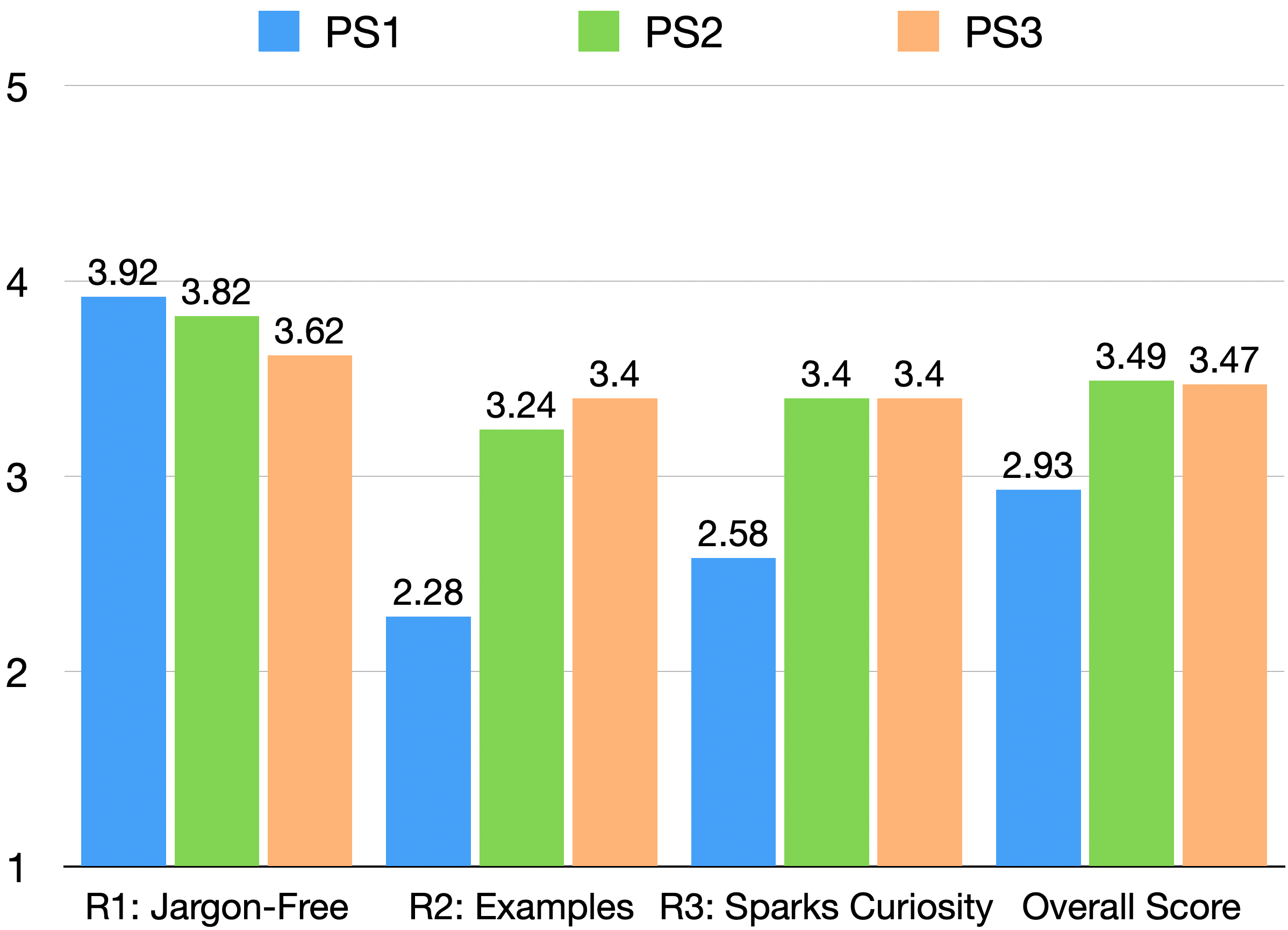}
  \caption{Average scores for each prompting strategy based on rubric performance}
  \label{fig:annotationresult}
\end{figure}

According to our annotation results (See Figure \ref{fig:annotationresult}), PS1 was the lowest-scoring strategy, with an average of 2.93 out of 5. PS2 and PS3 were only about half a point higher than PS1 at 3.49 and 3.47 out of 5, but about equal to each other. 
All three strategies performed pretty well at being jargon-free, even PS1. Seemingly, LLMs can follow the instruction to be jargon-free without examples. However, where PS1 struggled was in being relatable and sparking curiosity. Here, PS2 and PS3 performed 1 point better on relatability and almost 1 point better on curiosity. This indicated that the training examples in PS2 and PS3 did help LLMs ``learn'' how to write a more relatable hook with details.  

% \vspace{-5px}
% \vspace{-15px}
% \input{figures/a_deactivated_table_study1pvalue}

To answer our Hypothesis \#1, \textbf{PS3 and PS2 were similarly good, and both were better than PS1.} Specifically, PS3 was only significantly better than PS1 for R2 (p-value $<$ 0.001), R3 (p-value $<$ 0.001), and the overall performance (p-value $<$ 0.001). However, compared to PS2, PS3 performed similarly to PS2 in all categories. This was surprising because the average score of PS2 (SCORE - 3.49/5) left much room for improvement. We hoped the chaining in PS3 would improve the hook quality, but it did not. 

One reason for PS3's unideal performance was that, PS3 often included jargon and failed to be relatable, though PS3 provided more detailed experiences. For example, the lowest-scoring hook from PS3 on Table \ref{table:collectionbadgood}, we saw that, with a topic of Back End, it did not give out a more detailed experience than what PS2 usually had: ``\textit{my recent experiences with Amazon Web Services' Identity and Access Management feature...}'' It reflected a problem that PS3 often included details that were specific but not relatable and even contained jargon or unexplained terms like ``\textit{Amazon Web Services},'' ``\textit{Identity and Access Management},'' and ``\textit{bad end access}.'' Clearly, this experience was not relatable to general audiences, though it was detailed. Thus, for the lack of improvement from PS2 to PS3, we can see the lack of manual filtering of the specific. However, with humans in the loop, the process of picking better answers would help improve answers at every step and make the final results closer to the rubric. Thus, to understand whether human interventions help with the PS3, we conducted the following study.

\section{Study 2: User Study}

We conducted a user study to evaluate the effectiveness of our LLM-based Tweetorial solution for users with the need to communicate science to the general public. 

\subsubsection{System Description}

\begin{figure}[!h]
  % \centering
   % \vspace{-0.35cm}
  \hspace{-6px}
  \includegraphics[width=1.07\linewidth]{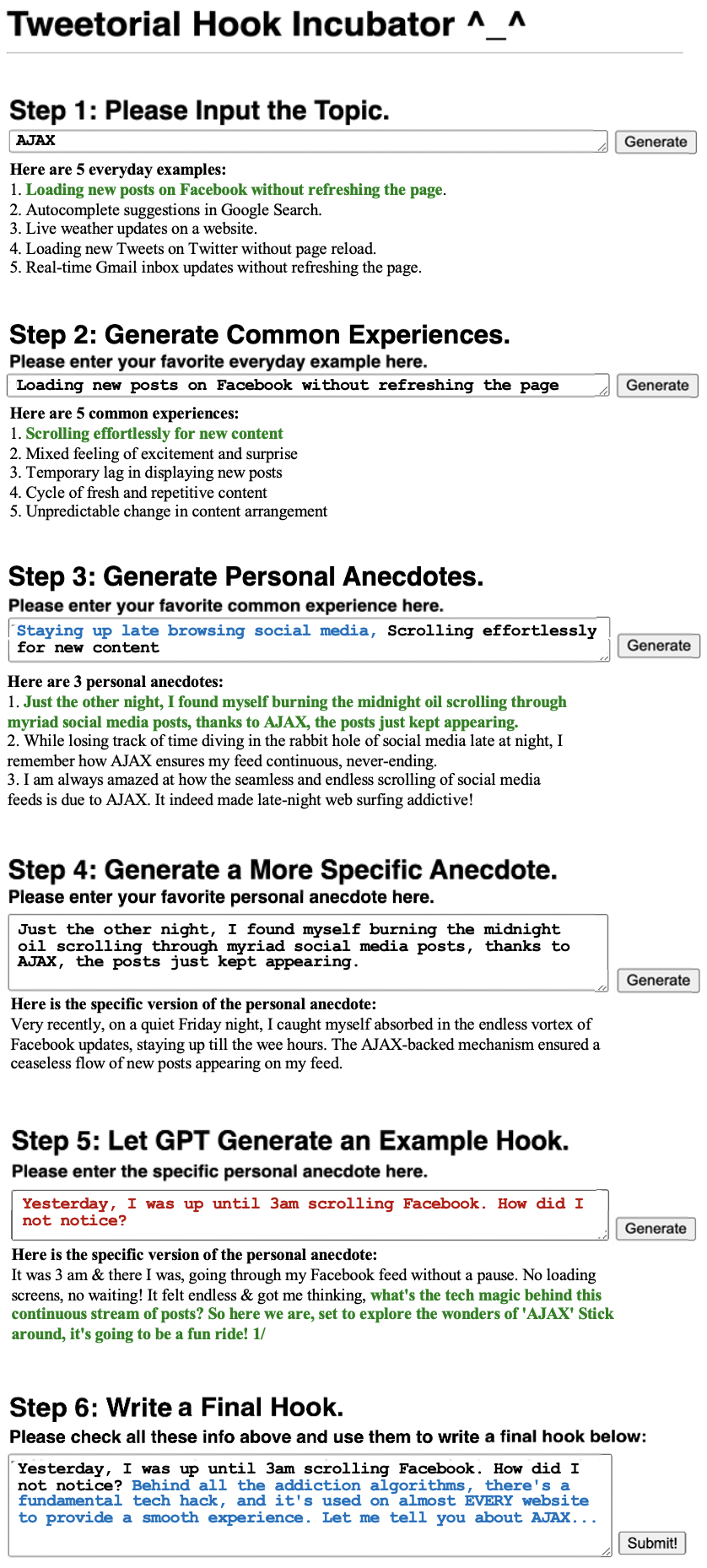}
  % \vspace{5px}
  \caption{An example of how users write a Tweetorial hook about ``\textit{AJAX}" with our tool, Tweetorial Hook Incubator. Users have the flexibility to navigate through the workflow: they can follow it sequentially from top to bottom, start from the middle steps, return to previous steps, or restart the workflow. They can \textcolor{OliveGreen}{\textbf{accept the LLM outputs}}, \textcolor{NavyBlue}{\textbf{adapt}}, \textcolor{Maroon}{\textbf{use their own responses}}, or regenerate. %We colored the input based on these conditions.
  }
  \label{fig:tool}
  \vspace{-0.95cm}
\end{figure}

We built an interactive web application using HTML, Python, Javascript, Flask, and the GPT-3 API to help users write engaging hooks for technical topics. The interface scaffolded the process of writing a hook into steps and used GPT to generate suggestions that the users can regenerate, modify, or accept before going to the next stage. The system and the workflow can be seen in Figure \ref{fig:tool}:

\begin{itemize}
    \item \underline{\textit{Step 1. Everyday Examples of Topic}}: Users input their topic, and the system generates five concrete everyday examples of that topic. For example, a user inputs the topic of ``\textit{AJAX}" from web programming, and the system generates five everyday examples such as ``\textit{autocomplete in Google Search}" and ``\textit{loading new posts on Facebook without refreshing the page}." The user picks an everyday example that is factually correct and relatable.
    % (see Figure \ref{fig:tool}).
    
    %In Figure \ref{fig:tool}, the user wanted to write a tweetorial about ``AJAX." They entered it as their topic in the input box and pressed the generate button. The user chose ``Loading new posts on Facebook without refreshing the page," because it was relevant and specific enough so people could understand it.
    
    \item \underline{\textit{Step 2. Common Experiences for an Everyday Example}}: Given an everyday example (from the previous step, or edited), 
    % Users input their everyday example (either from the previous step or their own answer), and 
    the system generates five common experiences people might have with that example. 
    For example, a common experience the system suggested relating to ``\textit{loading new posts on Facebook}"  is ``\textit{Scrolling effortlessly for new content}." The user likes the relatable feeling of ``\textit{scrolling effortlessly}" but wanted a more vivid experience that would resonate even more with users. Inspired by the system, the user wrote: 
     ``\textit{Staying up late browsing social media}." 
     % (see Figure \ref{fig:tool}).
    
    % In Figure \ref{fig:tool}, after the user copied their favorite everyday example from the first step, they pressed generate to find common experiences. The user liked the second common experience generated but still added a few more details to make it resonate more. Thus, the user moved forward with the experience of ``Scrolling effortlessly for new content" but added some detailed experiences, like ``Staying up late browsing social media."
    
    \item \underline{\textit{Step 3. Sample Personal Anecdote}}: 
    Given a common experience, the system generates three personal anecdotes and narratives. For example, 
    %the user combined the GPT-generated common experience and added some personal details to generate personal anecdotes. After generating, the user found that the 
    the system generates three sample anecdotes that rephrase the common experience in a first-person view.
    % about ``staying up late browsing social media effortlessly." 
The user liked the phrasing  ``\textit{just the other night, I found myself [scrolling Facebook]}" -  it aligned with their own experience and felt relatable. They didn't like some of the dated language (``\textit{burning the midnight oil}"), but they were willing to see a more specific version of the anecdote.
% (see Figure \ref{fig:tool}).
    
    \item \underline{\textit{Step 4. More Specific Personal Anecdote}}: 
    Given a short personal anecdote, the system generates a new version with more specific details. Here, it made ``\textit{Just the other night}" more specific by saying ``\textit{a quiet Friday night}". %Users input their favorite personal anecdote back to the system to add details to make it specific and vivid. In Figure \ref{fig:tool}, the user pressed 'generate' 
    Here, the details weren't correct and weren't particularly engaging. Thus, they rewrote their own anecdote by drawing from their personal experience, with similar types of specific language, but more succinct and authentic to their experience: ``\textit{Yesterday, I was up until 3 am scrolling Facebook}." 
    % (see Figure \ref{fig:tool}).
    
    \item \underline{\textit{Step 5. Sample Hook for a Specific Anecdote}}: Given a specific personal anecdote, the system generates an example hook based on all previous inputs. For example, the user liked their own personal anecdote as a specific and relatable example (R2), but 
    %after the user inputted their self-written specific anecdote and pressed 'generate,' they received a ready-to-use example hook generated by GPT. 
    the system generated a good way to spark curiosity (R3): ``\textit{What's the magic behind this continuous stream of posts?}" But the user adapted the language to be more emotionally heightened:
    \textit{``Behind all the addiction algorithms, there's a fundamental tech hack and it's used on almost EVERY website to provide a smooth experience.''}.
    
    % ready-to-use hook that connects all the details from previous steps, such as ``scrolling effortless," ``I was up 3 am scrolling Facebook." The user then finds the latter half particularly engaging, as it effectively hooks the audience with phrases like ``what's the tech magic," ``set to explore the wonders," and ``a fun ride." 
    % However, the user feels the first half is just a GPT rephrase of their personal anecdote, and its tone is a bit too dramatic. As a result, they choose to combine their original anecdote with the later half of the example hook (see Figure \ref{fig:tool}).
    
    \item \underline{\textit{Step 6. Final Hook}}: Users input their final hook into the text box. They can directly copy the LLM-generated hook from Step 5, they can adapt it (as seen in Figure \ref{fig:tool}), or write their own with inspiration from some of the ideas in previous steps. They click ``Submit'' when they are done.
    % or adjusting it accordingly. For example, as the user finds the GPT-generated example hook had some useful parts and others that needed adjustments, they modified it to construct their final hook. 
    % Then, they clicked 'Submit' to conclude a writing session about 'AJAX' (see Figure \ref{fig:tool}).
\end{itemize}

\subsubsection{Participants and Procedures}
We recruited ten participants from a local college student network and asked them to write Tweetorial hooks with and without our prototype in February 2023. The participants included six females and four males, with an average age of 20.1 years old. All ten users had expertise in computer science and familiarity with the particular topics we were asking them to write about. The study took around 1.5 hours, and they were paid \$30. 

Before the study, 
% they 
participants
first received a 10-minute introduction to Tweetorials and hooks. The introduction included explanations and examples of what constitutes a good hook. Then, they were asked to write hooks for six randomly chosen computer science topics from the list we used for the annotation study. The topics, in sequential order, were \textit{Front End}, \textit{Autocomplete}, \textit{Programming Language}, \textit{Net Neutrality}, \textit{Application Programming Interface (API)}, and \textit{Cybercrime}. For the hook-writing tasks, we asked each of them to write on three topics using the system and three without the system. The participants were randomly assigned to two groups, each consisting of five. Group 1 wrote with the system for the first, third, and fifth topics, and without the system for the others. Group 2 followed the opposite order. This approach ensured a fair comparison by evenly distributing the system use across all topics and participants. 

During each hook-writing task, we first provided the participants with the topics and informed them whether to use the system. Then, they had eight minutes to write a hook. During the session, users were informed that they could search for information online regardless of the conditions. After each hook writing task, we asked them to fill out a NASA Task Load Index (TLX) \cite{nasatlx} questionnaire to understand their mental load and experiences quantitatively. After finishing all six writing tasks, we started a 25-minute semi-structured interview to learn more about their experiences and hook writing process. 

In this study, we investigated the following hypothesis:

\textbf{Hypothesis \#2: Using the system reduces the mental load and increases the performance of writing hooks.} 

% \newpage

\subsection{Results}
\begin{figure}[!h]
  \centering
  \includegraphics[width=\linewidth]{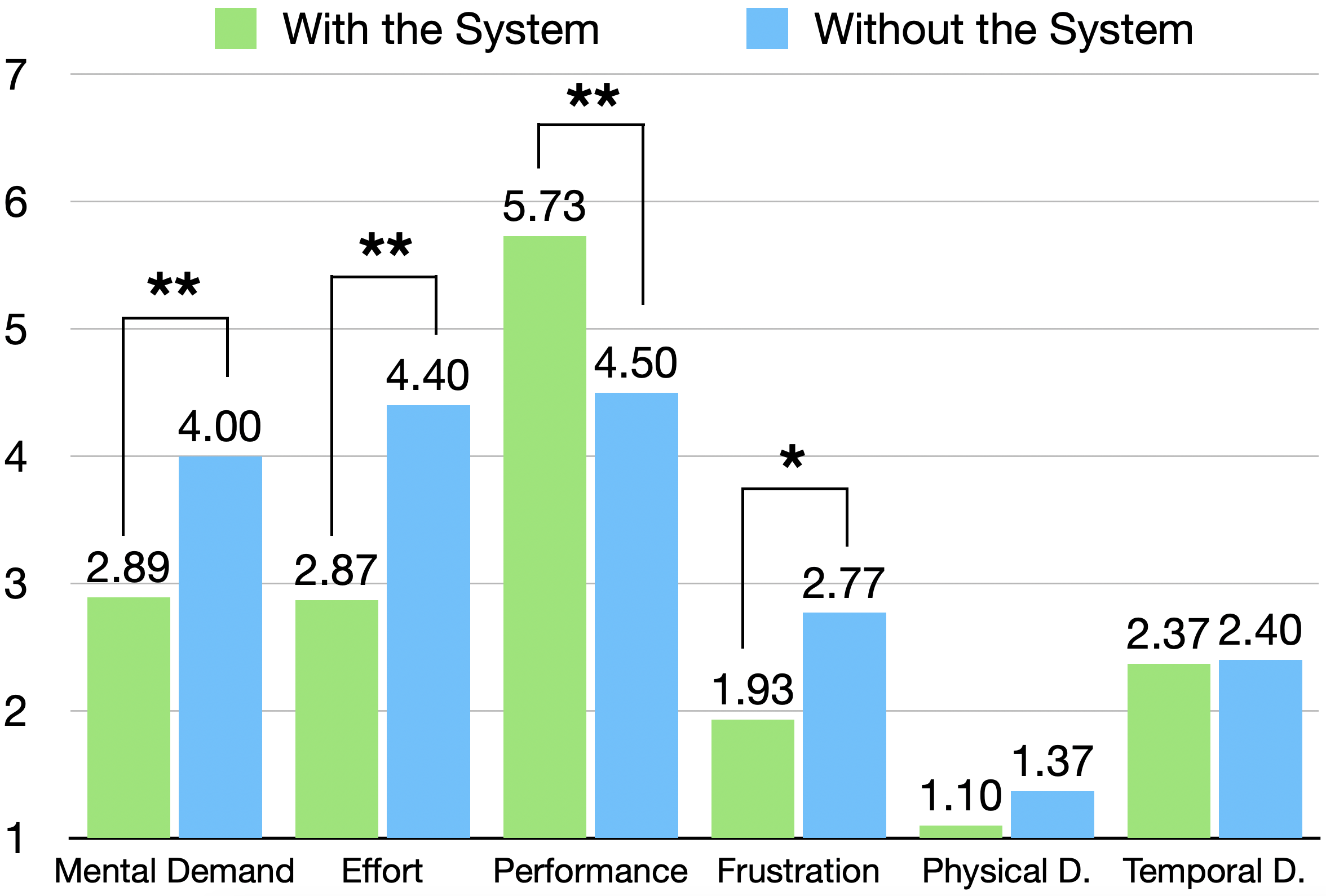}
\caption{User study TLX results (** indicates statistical significance at the p$<$.005 level, * indicates statistical significance at the p$<$.05 level)}
  \label{fig:tlx_results}
\end{figure}

The TLX results are visualized in Figure \ref{fig:tlx_results} and Table \ref{tab:userstudy_statistics}. As we split participants into two groups for randomization, they had good internal consistencies within each group, with Cronbach's Alphas of 0.78 and 0.85.

\vspace{-0.2cm}
% \vspace{-0.5cm}
\bgroup
\def\arraystretch{1.1}
\begin{table}[!h]
\centering
\small
\begin{tabular}{|l|l|l|l|} 
\hline
\textbf{TLX Dimension} & 
\textbf{With System} & 
\textbf{Without} &
\textbf{p-value}
  \\ 
\hline
Mental Demand & 
2.87 & 
4.00 &
\textbf{0.004}\textbf{**}
  \\ 
\hline
Effort  & 
2.87 & 
4.40 &
\textbf{0.002}\textbf{**}
  \\ 
\hline
Performance  & 
5.73 & 
4.50 &
\textbf{0.001}\textbf{**}
  \\ 
\hline
Frustration  & 
1.93 & 
2.77 &
\textbf{0.02}\textbf{*}
  \\ 
\hline
Physical Demand  & 
1.10 & 
1.37 &
0.08
  \\ 
\hline
Temporal Demand  & 
2.37 & 
2.40 &
0.598
  \\ 
\hline
\end{tabular}
\caption{User study TLX results and p-values for Wilcoxon tests (** indicates statistical significance at the p$<$.005 level, * indicates statistical significance at the p$<$.05 level)}
\label{tab:userstudy_statistics}
\end{table}
\egroup

\bgroup
\def\arraystretch{1.15}
\begin{table*}[!ht]
\centering
% \vspace{-0.25cm}
\small
\begin{tabular}{|p{1.8cm}|p{6.7cm}|p{7.8cm}|} 
\hline
 \textbf{Topic} &\textbf{Without the System} &
  \textbf{With the System}  \\ 
  \hline

% Application \newline Programming \newline Interface \newline (API)                         &
% Have you heard about the huge Oracle and Google lawsuit but had no idea what it was about? What the hell even is an API, and why is it so important that they can't be copyrighted? A guide for Supreme Court Justices (and you)   &
% Ever wonder why your Spotify Wrapped is so fun? How do they know which artists and songs to highlight and recommend? Find out how Spotify's developer tools can help analyze user listening history and trends to make tailored music content in this thread:\\ 
% \hline

% Cybercrime  & \colorbox{yellow}{These days computers are a huge part of our lives-} what illegal activities could be going on within our computers? In this thread, we will be exploring cybercrime, and what this could mean for our online safety. 1/  & \colorbox{yellow}{Have you ever received a call out of the blue from someone} \colorbox{yellow}{claiming to be from your bank, asking for your personal} \colorbox{yellow}{information?} After this happened to me recently, I wondered what other kinds of cybercrime exist and how someone like me can protect themselves? Here's what I found out:  \\
\setlength{\fboxsep}{0pt}

Cybercrime  & \colorbox{lightyellow}{\strut These days computers are a huge part of our lives-}\vspace{-0.3em} what illegal activities could be going on within our computers? In this thread, we will be exploring cybercrime, and what this could mean for our online safety. 1/  & \colorbox{lightyellow}{\strut Have you ever received a call out of the blue from someone}\vspace{-0.75em} \colorbox{lightyellow}{\strut claiming to be from your bank, asking for your personal}\vspace{-0.7em}  \colorbox{lightyellow}{\strut information?}\vspace{-0.3em} After this happened to me recently, I wondered what other kinds of cybercrime exist and how someone like me can protect themselves? Here's what I found out:\\
\hline
\end{tabular}
\caption{Collection of hooks generated in both the ``with-system" and ``without-system" conditions from the user study. Both examples are jargon-free (R1) and contain specific and relatable examples (R2, highlighted). Notably, the example from the ``with-system" condition includes more specific details to resonate with users.}
\label{tab:user}
% \vspace{-0.4cm}
\end{table*}
\egroup

\noindent \textbf{1. Less Mentally Demanding} 
The TLX scores indicated that writing hooks was less mentally demanding with the system (SCORE - 2.87/7) than without it (SCORE - 4.00/7, p-value = 0.004). %(See Figure \ref{fig:tlx_results}.)
All ten users expressed that without the system, it was hard to find concrete and specific examples of abstract topics. Under that condition, many users did their own brainstorming, often trying to think of their own experiences with the topic and attempting to recall tangible details and emotions about it before they were able to start writing (P8).
Five users said that even if they did come up with a few examples, it was challenging to narrow them down to one to fit the criteria: relevant, relatable, and interesting enough to make them keep reading (P1, P2, P5, P7, P8). 

All ten users expressed the ease of using the system to help simplify language into digestible terms that more people can understand. P4 shared it is easier to brainstorm a lot of ideas, and it helped open horizons and applications, but they still ended up choosing one that resonated the most. P1, P2, P5, P7, P8, and P10 mentioned that the workflow was straightforward, easy, clear, and simple to use, easing mental burdens during the hook writing process. All ten users said they would use this tool in the future.

\medskip
\noindent \textbf{2. Less Effort}
The TLX scores indicated that writing hooks required less effort with the system (SCORE - 2.87/7) than without it (SCORE - 4.40/7, p-value = 0.002). 
Under the without system condition, seven users spent a lot of effort searching the Internet to find examples without much success. 
Even though there were some examples on Google, it was hard and time-consuming for users to find them. P2 and P5 shared that Google felt like an “\textit{ocean of information}.” They had to spend a long time searching: skimming through the titles, avoiding getting technical information, and clicking on it to understand the material first and then adapting it to their own work. They needed to put down three to five search queries on Google to find the results they wanted. For example, P9 used “\textit{net neutrality examples}” and “\textit{net neutrality in simple terms}”; P2 used “\textit{examples of APIs we use in our everyday lives}'', “\textit{define programming language in a fun way},” “\textit{explain the term front end for a 5-year-old}” and “\textit{what is the front end for dummies}”; P7 used “\textit{examples of popularly used APIs}” and “\textit{how to talk about programming languages in layman's terms}.” Trying different terms took a long time and effort (P8) and often ended in failure (P1).  

In contrast, P1, P2, P4, P6, and P7 all mentioned that the with-system experience was just effortless: ``\textit{easy to generate and regenerate}'', ``\textit{easy to find strong ideas}'', and easily ``\textit{reminded me of what I already knew}''. P8 shared that the writing workflow was seamless, enabling them to complete the hook writing process by following the steps without searching on Google. In total, eight of the ten participants finished the with-system writing process without Google.

\medskip
\noindent \textbf{3. Better Performance}
The TLX scores indicated that users achieved better performance writing hooks with the system (SCORE - 5.73/7) than without it (SCORE - 4.50/7, p-value = 0.001). Also, from the table \ref{tab:user}, users felt more confident and satisfied with the results they obtained from the system when using LLMs, as they believed that the process involved fewer personal biases and LLMs had more knowledge about real common experiences. For instance, P8 mentioned that they believed the common experiences generated by LLMs were meant to be more familiar and relatable to the general public. In comparison, they reported concerns that the experiences they came up with on their own or from Google were not common enough and biased toward their personal background. Similarly, P2, P4, and P7 shared that they experienced these implicit biases and received fewer affirmations while trying to write a hook without LLMs, as they trusted LLMs more.

\medskip
\noindent \textbf{4. Users Edit LLM Hooks 
% for the 
to Meet
Requirements}
In Step 5 of the system, users were presented with a hook written by the LLM based on their responses to Steps 1-4. 
All ten users expressed that the LLM-generated hooks are good and useful, while six of them expressed the need to edit the LLM-generated hooks to make them more relatable and engaging.
When asked to make a quick comparison between their edited version and the LLM-generated ones, all these six writers shared that their edits were necessary and helped elevate the quality of the hooks. 

Responding to R1 (being jargon-free), P1, P8, and P10 shared that they still found jargon inside the LLM-generated hooks. Thus, they removed the unexplained terminology or hard-to-understand acronyms. For example, P10 replaced the acronym of ``\textit{ISPs}'' from the LLM-outputted hook with ``\textit{Internet Service Providers}.'' They had concerns that the system might 
% oversee 
overlook requirements after chaining too much stuff. Also, they edited the hook for conciseness by cutting off extra questions and wordy introductions.

For R2 (including relatable and specific examples), several writers said that the LLM output felt robotic and rigid, thus making it less engaging (P1, P2, P5). For example, P1 mentioned that when they read the LLM-generated hook, they felt it would not interest readers. Also, P2 shared that the first sentences in many LLM-generated hooks felt like news headlines, which read like some emotionless statements. Thus, they edited the tone to become funnier and more personable. Also, P10 shared that they changed the time-related examples inside the hook as LLMs sometimes lacked updated information. Hence, they replaced the LLM output with a more recent example. 

For R3 (sparking curiosity), P1 and P4 shared that they know what makes a tweet go viral and get clicks from their past Twitter experiences: using exaggeration, shock factors, and potentially misleading information. Then, P4 prepended \textit{``Apparently we're gonna lose \$10.5 trillion to criminals over the internet by 2025. Isn't that horrendous?}'' to the LLM-generated hook on cybercrime. They believed the addition of surprising data would attract readers more.

\medskip
\noindent \textbf{5. Users Edit LLM Hooks for Personal Style} 
Users also edited the LLM-generated hooks to make them fit more according to their writing styles and favorite examples so they felt more connected and related to their hooks. 
For example, P10 shared that they wanted to use the exact syntax they used daily in this hook. So they changed a lot of word-level choices like from “\textit{Do you know what}” to “\textit{Have you ever heard.}” P10 also shared that they intentionally deleted words like ``exactly'' and split the two questions which were originally in one sentence into two separate short ones. From this, P10 shared that it made them feel that the hook sounded like themselves or their friends by referring to their usual language choices. Also, P1 and P10 edited all of the LLM-generated hooks when they reached Step 6, even though they stated they were already highly satisfied with them. They still expressed wanting to embed more of their styles inside the hook. P4 suggested that making these changes helped maintain their own voice, and P6 specifically added several hashtags and emojis as they liked them. According to P8, engaging in the final editing of the hooks helped them feel a greater sense of agency and ownership over them. This was because they perceived the final product as being more original after undergoing the editing process. P8 specifically mentioned while editing, they shifted from the role of ``creator'' to the ``first reader'' of the hooks. By doing so, they gained a more objective and distant view of their writings.

\section{Discussion and Future Works}
In this paper, we demonstrate that LLMs can help contextualize technical information into relatable and engaging hooks. We scaffold the complex Tweetorial hook writing process by prompting LLMs for everyday examples, common experiences, and specific anecdotes. This scaffolding approach~\cite{scaffolding} helps STEM experts effectively communicate science to non-technical audiences.
% , especially those who lack experience or understanding of audience perspectives from diverse demographics and backgrounds. 
In the future, it is possible that
similar tools could be built for other groups of experts, such as helping journalists reach younger audiences, helping medical professionals explain procedures to patients, or helping public service organizations spread messages to under-served communities.

However, LLMs are far from perfect and user interaction is essential to producing successful hooks. 
LLMs sometimes provide inaccurate examples for a topic and sometimes suggest experiences that a non-technical audience would not relate to, such as building a website or buying something on the dark web. 
Ultimately, the expert must decide whether the suggestions are correct and appropriate, and they cannot just ``trust the machine.'' Experts have the ability to judge whether the examples of the technology are correct (such as verifying that Spotify Wrapped does indeed use an API), but they might not understand non-technical audiences well enough to evaluate whether the suggested experiences resonate with them (such as being aware of a lawsuit between Oracle and Google). If an expert is unsure whether something would resonate with the public, they should ask members of their audience. One feature that could be built into such as system is to get human judgments from an online marketplace to provide audience feedback on demand.

Although LLMs have a wealth of information, they do contain biases and not all viewpoints are equally represented. For explaining science to the general public, the biases in the current LLMs like GPT-3 and GPT-4 are probably not problematic. However, if the intended audience were a more specific demographic, LLMs might not suggest examples and experiences that resonate with them. People of different ages, cultural backgrounds, education levels, language abilities, and geographic locations communicate very differently. For example, an experience about using a laptop might not resonate with a low-income student who cannot afford a laptop and does all of their computing from a phone. Currently, LLMs mostly echo dominant perspectives, but it could be powerful to train LLMs to elevate the voices of non-dominant groups as a means to bridge the gap, better support the communities, and promote inclusivity. 

% In the case of STEM experts explaining concepts to a general audience, many of the STEM experts can probably make that judgment by thinking of a family member who is not in STEM. However, for other populations, like under-served communities, it might be harder for an expert to judge whether the material would resonate. More work should be done to see to what degree LLMs can generate suggestions for populations with different demographics: ages, geographic areas, cultural backgrounds, education levels, etc.  Although, getting insights and feedback from real members of the target audience would be the best solution. 

In the study, users stated that it was important to them that their final hooks reflected their own personal style and creativity. 
This is in line with previous work on the social dynamics of AI co-creative systems~\cite{socialdynamics} which has shown that when working with LLMs, writers care deeply about preserving their \textit{intent} and the \textit{authenticity} of their writing. To further enable this, 
some users suggested future versions of the system where writers can feed their hooks back to their system to ``keep'' their style for future generations, or add a “temperature” parameter to control the specificity of contextualized examples. These features can provide a range of agency when co-creating hooks with LLMs, thus aligning with the future vision of designing more user-focused interactive creativity support tools. These designs can empower users in their content creation process by fostering a sense of ownership and creative expression.

\section{Conclusion}
This paper explores integrating generative AI into the hook writing process for Tweetorials, a science communication method that motivates science through relatable examples and experiences. 
Our prompting engineering study suggests that including examples of good hooks in the prompt helped LLMs generate better hooks, but there is still a need for humans in the loop.
To help experts write hooks, we built an LLM-based workflow that scaffolds the process: given a topic, the system suggests everyday examples of the topic, and the user can accept a machine suggestion, edit a machine suggestion, request more suggestions, or write their own. Based on the everyday example selected by the user, the system suggests common experiences. The user can again accept, edit, regenerate, or write their own. Based on the common experience selected, the system suggests a personal anecdote and can make the anecdote more specific while the user may edit these as well. Finally, the system produces an example hook that users can accept as is, or reference when finalizing their hook. Our user study shows this scaffolding  greatly reduces the cognitive load of writing hooks. Also, as the outputs are editable at every stage, the hooks still convey the writer's authentic style, voice, and experiences. 
% common experiences with the an example, personal anedcotes based on the experience, a more specfic anecdote, and finally an example hook. At each stage the user can accept a machine suggestion, edit a machine suggestion, request more suggestions, or write their own answer. 
% first the user inputs their topic, then the LLM suggests five everyday examples. the user can pick one, edits  by suggesting everyday examples, 
% LLMs an interactive workflow 
% the best practice of writing a hook using LLMs should include chaining of prompts: examples of hooks and contextual user details that are specific and relatable to users. 

% and produced a more engaging and relatable hook to motivate readers. 
% The edits made by users to the LLM outputs contribute to our understanding of how users make decisions when utilizing co-creation tools.

\section{Author Contributions}
TL finalized the prompt engineering works, built the system, led the annotation and user study, analyzed the results, and wrote the paper. DZ led the early prompt engineering and task understanding and contributed to the system and data collection. GL, BT, SM, and KS assisted with the early task understanding, data collection, and analysis. SW, KG, and LC provided overall guidance on the project, helped shape the two studies, and contributed to the writing.

\section{Acknowledgments}
This work has been supported by NSF-IIS-2129020 and NSF-EAR-2121649. 

% \newpage
% \newpage
\bibliographystyle{iccc}
\bibliography{iccc}

\begin{figure*}
% \vspace{-0.2cm}
\centering
{\section{Appendix}
\label{appendix:A}}
\vspace{0.1cm}
{\textit{Due to the page limits, a high-resolution source appendix is linked here: {\footnotesize \url{ https://tinyurl.com/tweetappendix}}}}
\end{figure*}

\begin{figure*}[!ht]
  \centering
  \includegraphics[width=\linewidth]{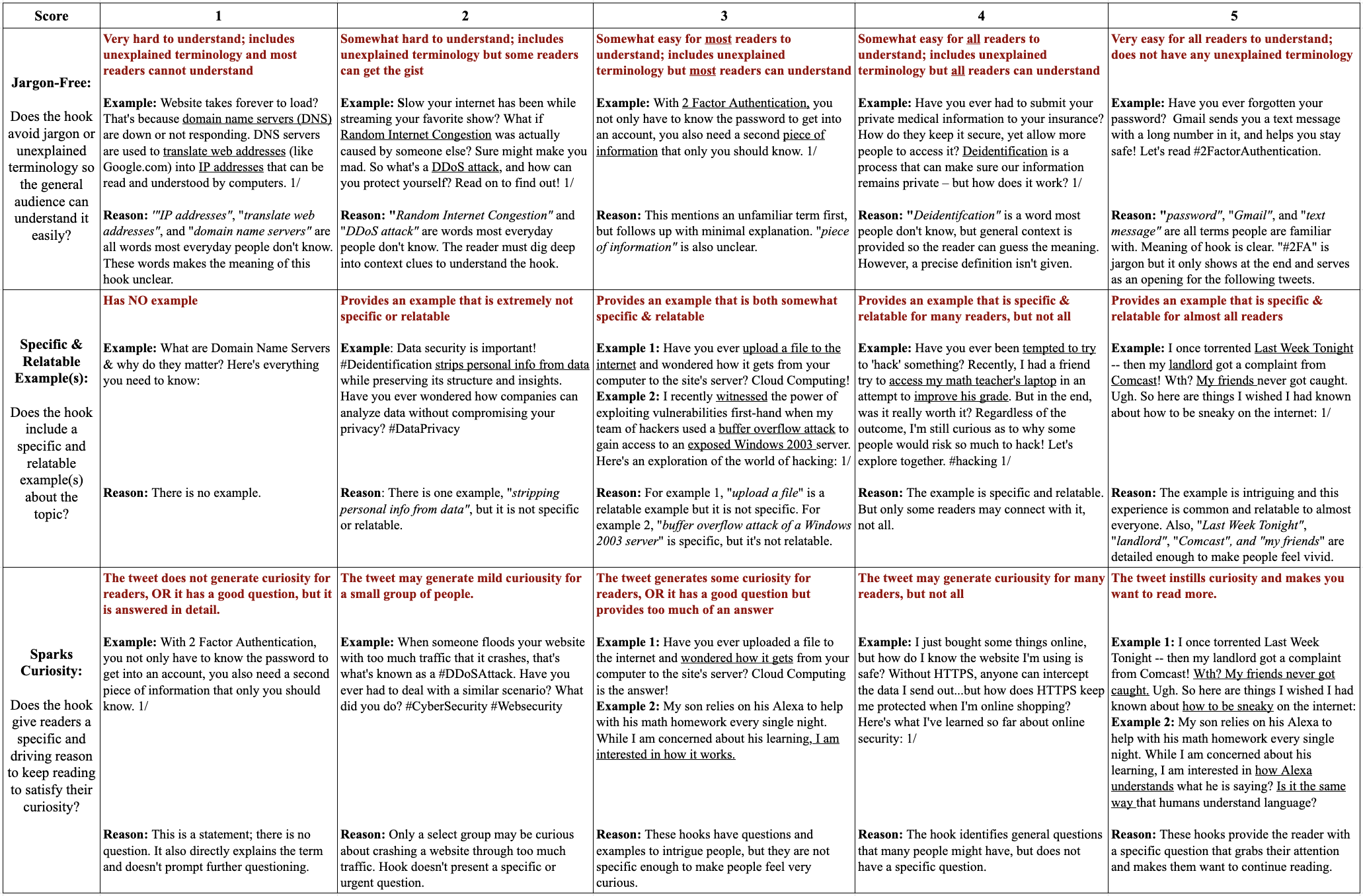}
\caption{A five-point scale rubric for annotators in the prompt engineering study}
\end{figure*}

\bgroup
\def\arraystretch{1.1}
\begin{table*}[!ht]
  \centering
  \tiny
  \begin{tabular}{|l|l|l|l|l|l|l|l|l|l|}
\hline
Ransomware&
Database&
Browser Hijacking&
Machine Learning&
\textbf{API*}&
Patch&
White Hat&
\textbf{Programming Language*}&
Trojan&
Ad Blocking
  \\ 
\hline
\textbf{Front End*}&
Peer-To-Peer&
\textbf{Net Neutrality*}&
Internet Service Provider&
Tor&
Black Hat&
BitTorrent&
Secure Socket Layer&
\textbf{Cybercrime*}&
Troll
  \\ 
\hline
Back End&
Browser&
Breach Notification&
Access Control List&
Dark Web&
OAuth&
Security Key&
\textbf{Autocomplete*}&
Backdoor&
Spoofing
    \\ 
\hline
\end{tabular}

\caption{30 topic inputs for the prompt engineering study (*asterisked are the six topics randomly chosen for the user study)} %The final score is averaged by ({\small Score\textsubscript{R1}, Score\textsubscript{R2}, Score\textsubscript{R3}}).}
\label{table:topics}

\end{table*}
\egroup

\bgroup
\def\arraystretch{1.1}
\begin{table*}[!ht]
  \centering
  \tiny
  \begin{tabular}{|p{0.65cm}|p{4.7cm}|p{5.4cm}|p{5.2cm}|}
\hline
\textbf{Strategy } & \textbf{Lowest-scoring hook}	&      
\textbf{Lowest-scoring hook with same topic as highest-scoring hook}       & \textbf{Highest-scoring hook}                                                     \\ 
\hline
\textbf{PS1 }& \textbf{Application Programming Interface (API)}
\newline
A \#programming interface that lets you calculate prices, map distances, and get stock quotes automatically? What else can an \#API do and how can it help me?\newline \newline \textbf{Average Score: 1.89 }                                                                                                                                                                                   & \textbf{Browser Hijacking}\newline A hacker could take control of your online activity and webpages without you realizing it! Ever heard of browser hijacking? Have you taken the necessary steps to protect yourself? \#browserhijacking \#cybersecurity \#hackerrisk\newline  \textbf{Average Score: 2.89}                                                                                                                                                                      & \textbf{Browser Hijacking}\newline 
Beware of browser hijacking! Ever gone to a website you didn't type in and been redirected somewhere else? That's browser hijacking! It can be used for malicious purposes - so how can you stay safe? \#browserhijacking \#securty \newline \textbf{Average Score: 4.00 }                                                                                                                               \\ 
\hline
\textbf{PS2} & \textbf{The Onion Router (TOR)} \newline Ever hear about the deep web and the black markets? Well, TOR is the gateway to them! What's required to risk taking a journey into the unknown? A thread on TOR and its implications: \newline\newline\newline\textbf{Average Score: 1.67 }                                                                                                                                                                                                      & \textbf{Internet Service Provider} \newline Have you ever wondered why you get more slow internet in your bedroom than at the dining table? What is it that your internet service provider does differently to affect your internet connection? Let's look into it: 1/ \newline\newline\newline \textbf{Average Score: 3.44                                                    }                                                                                                                       & \textbf{Internet Service Provider} \newline Have you ever had an internet outage at the exact moment when you need to send an assignment or video chat with friends? Who's responsible for this? How are we connected to the internet? I'm about to explore what happens behind the scenes of the internet. Let's start by understanding the main characters involved - the Internet Service Providers. Ready? 
\newline\textbf{ Average Score: 4.55 }
\\
\hline
\textbf{PS3} & \textbf{Back End} \newline Have you ever wanted to keep your data secure and away from prying eyes? You might not realize it, but controlling access to your accounts is just as important as data privacy. But how can you do this? My recent experience with Amazon Web Services' Identity and Access Management feature has shown me the power of properly managing back end access. Let's discover more! 1/\newline \textbf{Average Score: 2.11}  & 

\textbf{Internet Service Provider} \newline Have you ever wondered why an Internet Service Provider (ISP) could make promises that never turn out to be true? I recently learned the hard-way when my streaming experience got slower even though I had chosen a top-tier plan from a big-name provider. Here's an exploration of how ISP's work and why you should take their promises with a grain of salt. \#ISP \#Technology \#Internet \#Promises 1/ \newline \textbf{Average Score: 3.55}   & 

\textbf{Internet Service Provider} \newline I used to think living in the suburbs would mean better internet connection. But my Verizon Fios service was so unreliable and slow that I felt like I was back in the dark ages! What can we do to better understand the nature of internet service providers, and how can they provide truly reliable service? Here's the story: \newline\newline \textbf{Average Score: 4.78 }
\\
\hline
\end{tabular}

\caption{Collection of good and bad hooks from the prompt engineering study} %The final score is averaged by ({\small Score\textsubscript{R1}, Score\textsubscript{R2}, Score\textsubscript{R3}}).}
\label{table:collectionbadgood}

\end{table*}
\egroup

\end{document}